**Title:** Computational synthesis of locomotive soft robots by topology optimization
short title: Topology optimized locomotive soft robots

**Authors:**
Hiroki Kobayashi,[1] Farzad Gholami,[2] S. Macrae Montgomery,[2] Masato Tanaka,[3,1] Liang Yue,[2] Changyoung Yuhn,[1] Yuki Sato,[1] Atsushi Kawamoto,[1] H. Jerry Qi,[2] Tsuyoshi Nomura[1*]

**Affiliations:**
[1] Toyota Central R&D Labs., Inc.; Bunkyo-ku, Tokyo 112-0004, Japan.
[2] The George W. Woodruff School of Mechanical Engineering, Georgia Institute of Technology; Atlanta, GA 30332, USA.
[3] Toyota Research Institute of North America, Toyota Motor North America; Ann Arbor, MI 48105, USA.

*Corresponding author. Email: nomu2@mosk.tytlabs.co.jp

**Abstract:**
Locomotive soft robots (SoRos) have gained prominence due to their adaptability. Traditional locomotive SoRo design is based on limb structures inspired by biological organisms and requires human intervention. Evolutionary robotics, designed using evolutionary algorithms (EAs), have shown potential for automatic design. However, EA-based methods face the challenge of high computational cost when considering multiphysics in locomotion, including materials, actuations, and interactions with environments. Here, we present a design approach for pneumatic SoRos that integrates gradient-based topology optimization with multiphysics material point method (MPM) simulations. This approach starts with a simple initial shape (a cube with a central cavity). The topology optimization with MPM then automatically and iteratively designs the SoRo shape. We design two SoRos, one for walking and one for climbing. These SoRos are 3D printed and exhibit the same locomotion features as in the simulations. This study presents an efficient strategy for designing SoRos, demonstrating that a purely mathematical process can produce limb-like structures seen in biological organisms.

**Teaser:** Body shape of pneumatic soft robots is synthesized via topology optimization without prior knowledge of the legs and arms.

**Main Text:**

**INTRODUCTION**

Locomotive soft robots (SoRos) have drawn prominent research attention in recent years due to their advantages of adapting to environments (*1*, *2*) and compatibility with humans (*3*). Due to their inherent material softness, SoRos can assume a high degree of freedom in their states (postures) compared with their rigid-bodied counterparts (*3*), enabling them to realize various modes of locomotion observed in biological organisms such as crawling (*4*–*10*), walking (*11*, *12*), climbing (*13*, *14*), swimming (*15*–*19*), and spinning (*20*). These modes of locomotion are often coupled with the shape of the SoRos (*21*). Consequently, various types of shapes inspired by biological organisms are employed, such as quadruped (*5*, *10*–*12*), caterpillar (*8*, *9*, *22*), snake (*1*), fish (*10*, *15*, *19*), and octopus (*16*, *23*). For instance, SoRos mimicking quadrupedal animals utilize four limbs for locomotion, stepping forward by moving one or more limbs at a time in a coordinated sequence, while supporting their body weight with the other limbs. Snakes-like SoRos have a cylindrical shape and use lateral undulation to propel themselves forward. Therefore, body shape plays an important role for realizing the prescribed locomotion.

However, designing body shape for specific locomotion modes is intrinsically challenging because of the high degrees of freedom in SoRos' postures (or states) (*24*), which have complex interactions with the surrounding environment. Many current designs of SoRo shapes mainly rely on the designers' knowledge and experience inspired by biological organisms, which acquired their sophisticated body shapes for efficient locomotion through million-year evolutionary processes. Because of differences in materials used, actuation mechanisms, and environments, bioinspired shapes (*9*, *16*, *19*) generally serve as the starting point and the final shape of the SoRos should be refined. To aid in the refining process, simulators are studied to reduce effort during prototyping and testing (*25*–*27*). Although simulators can accelerate the design process by automating performance evaluation, human interventions remain in determining the overall layout and shape of SoRos.

Another intriguing way to design the shape of locomotive SoRos is the evolutionary robotics approach, which mimics the evolution process in nature and uses evolutionary algorithms (EAs). Such attempts were made in a study by Sims in 1994 (*28*). In the EA-based design approach, a computational model is usually used to conduct forward prediction based on the designs provided by EAs. In Sims' pioneering work, moving objects were computationally created by combining the primitive shapes of rigid parts. A more complex design for actuating soft bodies was explored (*29*, *30*) by introducing a material distribution for structural representation. For the EAs, a structure is typically represented by the selection of voids, solids, and actuators in a voxel. Bhatia et al. presented a variety of tasks, including object carrying, climbing, and jumping on undulating terrain (*31*). An approach targeting specific applications, that is, pneumatic soft actuators, has also been proposed using a network-based representation (*32*). Furthermore, an EA has been applied to an organoid in vitro, using cell-based construction (*33*).

Since EAs mimic natural evolutionary mechanisms, it can be assumed that they are suited to generate shapes in nature through convergent evolution, ideally. However, EA-based SoRo design has two drawbacks. First, although numerous computational studies have been conducted on SoRos (*29*–*31*, *34*, *35*), few have been experimentally verified (*29*), which is essential for practical SoRo development. Second, although EAs can be combined with various forward prediction methods and dimensionality reduction methods for design variables (*36*), they not only need to evaluate a large population (or design) in each generation to suggest the next one but also require many generations (*37*). In previous studies on EA for actuating soft bodies (*30*, *31*, *34*), the total number of required simulations was substantial: 30,000 simulations (30 individuals per population, 1,000 generations) in (*30*), 25,000 simulations (25 individuals per population, 1,000 generations) in (*31*), and 600,000 simulations (300 individuals per population, 2,000 generations) in (*34*). This large number population size requires that each design candidate must be evaluated at a low computational cost. However, as we consider the designs using real engineering materials with specific actuation mechanisms and realistic interactions with the environment, more sophisticated computational models are needed and the performance evaluation of each candidate becomes more challenging and more time-consuming. To overcome these issues, an efficient computational paradigm is required to replace EAs.

Topology optimization (*38*) is a powerful approach for structural synthesis which formulates the structural optimization problem as a material distribution problem, allowing for the spatial allocation of either voids or materials. This approach typically employs gradient-based optimization for efficiency. In gradient-based topology optimization, the intermediate state of material is introduced for continuous transition from void to material to derive the gradient of an objective function for design updates. Topology optimization offers a high degree of freedom due to its efficiency and continuous design variables, allowing it to accommodate topological changes in the structure. It has been studied for complex physical phenomena, such as hyperelasticity with geometric nonlinearity (*39–41*), contacts (*42–44*), fluid–structure interactions (*45–47*), transient problems (*48–50*), and 4D printing (*51*, *52*). Pneumatic soft grippers have also been designed using topology optimization (*53–56*). However, few studies have considered all of these.

Recently, a topology optimization method based on the material point method (MPM) was proposed for soft bodies (*57*). The MPM is a numerical approach that simulates the behavior of continuum materials, including soft bodies and fluids, using a combination of particles and grids representation (*58*, *59*). MPM-based topology optimization was extended to include the design for the layout of multiple actuators and their respective time-series actuation (*60*). The actuator layout problem has also been studied in (*61*), considering soft and stiff material distribution. Aside from topology optimization, research has been conducted using the MPM to optimize SoRos moving over various terrains (*62*). Since these studies (*57*, *60*, *62*) were not intended to be applied to actual SoRos, they did not model practical actuators such as pneumatic actuators. Instead, it is assumed that the robot's soft body itself expands and contracts, which is difficult to realize in the real world. In addition, the viscosity of soft material, which affects the dynamic characteristics in locomotion, has not been considered. A recent study presented a structural optimization method of a pneumatic walking robot using MPM (*63*). Although the optimized design was fabricated and tested, the simulation was based on a 2D model, which did not consider complex deformations caused by air pressure. A comprehensive framework that can provide implementable designs for SoRos has not been sufficiently explored. Therefore, we focus on SoRos equipped with pneumatic actuators, which are the most widely used and easy-to-implement soft actuators in practice (*64*).

In this study, we proposed a gradient-based topology optimization method for designing pneumatically actuated SoRos that can function in the real world. The structure was represented by a spatial density distribution which continuously transitions from void to material. The SoRo shape was synthesized through iterative updates in a gradient-based optimization without an initial guess of the layout and shape of limbs (**Fig. 1A.**). The simulation was based on the MPM, incorporating visco-hyperelasticity to reproduce the material behavior and solid/fluid modeling to capture the interaction between soft bodies and air in the chamber. Using the proposed method, optimized SoRos can be fabricated for various locomotion tasks by 3D printing. We demonstrated the proposed method for two types of locomotion tasks: horizontal movement on a flat ground (walking) and vertical movement between two walls (climbing). For each task, the optimized shapes and their locomotion modes were studied by analyzing the temporal posture change by actuation. The gap between simulation and experiment was discussed by comparing the locomotion distance and the temporal trajectories of the parts. This research provides an efficient strategy for designing shapes of pneumatic SoRos. It also suggests that a purely mathematical process can produce limb-like structures and movements that resemble those found in the locomotion of biological organisms.

# RESULTS

## Shape synthesis of pneumatic SoRos for locomotion

The shape synthesis process using topology optimization is illustrated in **Fig. 1**. We defined a cubic design domain wherein a material point could be spatially designated as either a void or solid (soft material). For gradient-based optimization, we introduced a fictitious density that can be continuously changed from 0 (void) to 1 (solid). A pneumatic actuator was placed at the center of the cube. The analysis domain of the SoRo was discretized into particles for the MPM simulation (**Fig. 1B.**). The fictitious density of each particle in the design domain was a design variable in the optimization problem. Air particles in the chamber were actuated by applying periodic pressure changes.

We performed the simulation based on MPM, which is a hybrid Lagrangian-Eulerian technique that offers the advantage of handling large deformations, contacts, and fluid–structure interactions. In the MPM, objects are represented by particles, and the governing equations are solved in a grid. Specifically, we used the moving least squares MPM (MLS-MPM, (*65*)). The movement of the SoRo was calculated in the forward simulation. Subsequently, a backward simulation (also known as backpropagation) was performed to derive the design sensitivity, which is the gradient of an objective function with respect to the design variable at the start of the simulation.

The simulation was implemented in a differentiable form such that automatic differentiation (AD, (*66*, *67*)) could be used. This differentiable physics simulation enabled the algorithmic derivation of design sensitivity. In addition, the simulation used parallel GPU computations. For such high-performance computing with AD, we employed Taichi (*68*), an open-source framework in Python. By fully exploiting its functionality, a high degree of design freedom can be achieved while considering the complex physical phenomena.

The design variables were updated using Adam optimizer (*69*) based on the design sensitivity to maximize the objective function. A constraint function was incorporated using an augmented Lagrangian method, based on a previous work (*60*). The optimized shape of the SoRo was synthesized by repeating the forward and backward simulations, and design variable updates (**Fig. 1C.**).

The particle densities obtained by optimization were post-processed into a volumetric field, and the interface between the void and solid, which was defined as 50% density, was extracted as the shape of the robot for 3D printing (**Fig. 1D.**). An air hole was added to connect the tube to apply pressure.

## Topology optimization of soft materials with fluid interaction

First, we define the domain of interest, $D$ is the design domain, $\Omega_w$ is the chamber wall domain, and $\Omega_{air}$ is the air domain in the chamber. The governing equations are the conservation laws of mass and momentum, as follows:

$$\frac{D\rho}{Dt} + \rho \nabla \cdot \mathbf{u} = 0, \tag{1}$$

$$\rho \frac{D\mathbf{u}}{Dt} + \nabla \cdot \boldsymbol{\sigma} + \rho \mathbf{g} = 0, \tag{2}$$

where $\rho$ is the density, $t$ is the time, $\mathbf{u}$ is the velocity, $\boldsymbol{\sigma}$ is the stress tensor, $\mathbf{g}$ is the gravity acceleration, and $D/Dt$ represents the material derivative. The difference in the formulation between the solid and fluid was owing to the Cauchy stress. First, we define the Cauchy stress in the solid domain $D \cup \Omega_w$ based on a generalized Maxwell model (*70*, *71*):

$$\boldsymbol{\sigma}_{\text{solid}} = \boldsymbol{\sigma}_0^H + g_\infty \text{dev}(\boldsymbol{\sigma}_0^D) + \text{dev}\left(\sum_{i=1}^{N_{\text{prony}}} \boldsymbol{\sigma}_i^D\right), \tag{3}$$

where $N_{\text{prony}}$ is number of terms in the Prony series, $g_\infty$ is the relative modulus of equilibrium element, $\boldsymbol{\sigma}_0^H$ and $\boldsymbol{\sigma}_0^D$ are the hydrostatic and deviatoric stresses of the equilibrium element, respectively, and $\boldsymbol{\sigma}_i^D$ is the deviatoric stress of each Maxwell element. The operator dev represents the deviatoric component of tensors, i.e., $\text{dev}(\cdot) = (\cdot) - ((1/3)(\cdot):\mathbf{I})\mathbf{I}$, where the operator : represents the inner product, and $\boldsymbol{I}$ represents the identity matrix. $\boldsymbol{\sigma}_0^H$ and $\boldsymbol{\sigma}_0^D$ are expressed as follows:

$$\boldsymbol{\sigma}_0^H = \frac{\partial \Psi^H}{\partial J} \mathbf{I}, \tag{4}$$

$$\boldsymbol{\sigma}_0^D = \frac{1}{J} \frac{\partial \Psi^D}{\partial \bar{\mathbf{F}}} \bar{\mathbf{F}}^T, \tag{5}$$

where $J$ is the determinant of deformation gradient and $\bar{\mathbf{F}} = J^{-1/3} \mathbf{F}$ is the isochoric part of the deformation gradient. We define $\Psi^H$ and $\Psi^D$ as a hydrostatic and a deviatoric part of a potential energy of the neo-Hookean hyperelastic material model, as follows:

$$\Psi^H(J) = \frac{1}{4} K\{(J-1)^2 + \log^2 J\}, \tag{6}$$

$$\Psi^D(\bar{\mathbf{F}}) = \frac{\mu}{2}(\text{tr}(\bar{\mathbf{F}}^T \bar{\mathbf{F}}) - 3), \tag{7}$$

where $K = (3\lambda + 2\mu)/3$ is the bulk modulus, and $\mu$ and $\lambda$ are the Lamé's constants. Further, the stress of the Maxwell element was defined as:

$$\boldsymbol{\sigma}_i^D = g_i \bar{\mathbf{F}} \int_{-\infty}^{t} \exp\left(-\frac{t-t'}{\tau_i}\right) \frac{d}{dt'}(\bar{\mathbf{F}}^{-1} \text{dev}(\boldsymbol{\sigma}_0^D) \bar{\mathbf{F}}^{-T}) dt' \bar{\mathbf{F}}^T, \tag{8}$$

where $g_i$ is the relative moduli and $\tau_i$ is the relaxation time.

Second, the Cauchy stress in the fluid domain $\Omega_{\text{air}}$ is defined as:

$$\boldsymbol{\sigma}_{\text{fluid}} = -p\mathbf{I} + \mu_v(\nabla \mathbf{u} + (\nabla \mathbf{u})^T) + \left(\zeta - \frac{2}{3}\mu_v\right)(\nabla \cdot \mathbf{u})\mathbf{I}, \tag{9}$$

$$p = -k(1-J) - p_{\text{act}}, \tag{10}$$

where $p$ is the pressure, $\mu_v$ is the shear viscosity coefficient, $\zeta$ is the volume viscosity coefficient, and $k$ is the bulk modulus of the fluid. $p_{\text{act}}$ is an offset that induces a pressure change representing actuation.

The boundary condition was set as a no-slip condition on the ground or walls. The no-slip condition was realized by applying $\mathbf{u} = 0$ if $\mathbf{u} \cdot \mathbf{n} < 0$ in the grid nodes, where $\mathbf{n}$ is the normal vector of the ground or wall surface. This process is a non-differentiable operation in automatic differentiation. However, MPM partially preserves the gradient information of particles in contact (*60*). This is owing to MPM's hybrid representation using particles and grids. In the MPM, information is transferred from particles to adjacent grids and back to the particles, partially reflecting the information of neighboring particles that are not in contact.

For the topology optimization, we employed the density method, which represents the structure shape by a fictitious density distribution $\gamma \in [0, 1]$. To simulate the physical phenomena that depend on the fictitious density, the physical property values in the solid domain were interpolated as follows:

$$\rho = \rho_0\left((1 - \epsilon)\gamma^3 + \epsilon\right), \tag{11a}$$
$$\lambda = \lambda_0\left((1 - \epsilon)\gamma^3 + \epsilon\right), \tag{11b}$$
$$\mu = \mu_0\left((1 - \epsilon)\gamma^3 + \epsilon\right), \tag{11c}$$

where $\rho_0$ is the density of the solid material, $\lambda_0$, and $\mu_0$ are the Lamé's constants of the solid material, respectively, and $\epsilon$ is an infinitesimal value to avoid numerical instability in the void domain. Note that there are several choices of the interpolation functions (Eqs. 11a–11c). In previous works of soft body topology optimization using MPM, a linear function (*57*, *60*) and a quartic function (*61*) were used to interpolate Young's modulus. When extracting surfaces for 3D printing at $\gamma = 0.5$ (50% density), it is necessary to ensure that structures contributing to the function are not lost. For this reason, the cubic function which has a small contribution from intermediate densities was adopted.

A filtering technique was used for topology optimization to ensure spatial smoothness of the structures. We employed a particle-based method presented in (*60*) with a cubic function profile. In addition, a normalized sigmoid function was employed as a projection scheme:

$$\gamma = \frac{1}{2}\left(\frac{\tanh(\beta\tilde{\phi})}{\tanh(\beta)} + 1\right), \tag{12}$$

where $\beta$ is a parameter that controls the shape of the function. Therefore, the design variable $\phi \in [-1, 1]$ was filtered to $\tilde{\phi} \in [-1, 1]$, and then projected to $\gamma \in [0, 1]$ by Eq. 12. This filtering and projection procedure was based on a previous study (*72*).

The objective of SoRos is to locomote in a prescribed environment. Therefore, we defined the objective function as the position change at the end of prescribed time interval in the designated direction $\mathbf{e}$ as follows:

$$\mathcal{L} = \left(\mathbf{x}_{g, t_{end}} - \mathbf{x}_{g, t_{start}}\right) \cdot \mathbf{e}, \tag{13}$$

where $\mathbf{x}_{g,T_{end}}$ and $\mathbf{x}_{g,T_{start}}$ are the center of gravity at the end of simulation and the start of actuation, respectively. The center of gravity can be derived from

$$\mathbf{x}_g = \frac{\int_D \rho \mathbf{x}\, d\mathbf{x}}{\int_D \rho\, d\mathbf{x}}. \tag{14}$$

In addition, a constraint function was introduced to reduce the intermediate density:

$$\mathcal{C} = \frac{\int_D \gamma(1 - \gamma)\, d\mathbf{x}}{\int_D d\mathbf{x}} \leq \mathcal{C}_{max}. \tag{15}$$

The function $\mathcal{C}$ should have a sufficiently small value of $\mathcal{C}_{max}$ to ensure a structure mostly composed of solids and voids. In this study, we empirically set $\mathcal{C}_{max}$ to 0.0125, which means the density field must not exceed 5% of maximum intermediate state (entirely filled with $\gamma = 0.5$).

The optimization problem was formulated as follows:

$$\max_{\phi \in [-1, 1]} \mathcal{L}, \tag{16a}$$
$$\text{subject to} \quad \mathcal{C} \leq \mathcal{C}_{max}, \tag{16b}$$

where the displacement field was governed by Eqs. 1 and 2. Since the optimizer Adam cannot directly handle constraint functions, we used the augmented Lagrangian method to replace Eqs. 16a and 16b with an unconstrained optimization problem, adding the Lagrangian term and the penalty term. The constraint (Eq. 16b) aimed to reduce the intermediate density and may lead to poor local optima. Therefore, we applied a minimal penalty at the beginning of the optimization and scheduled an increase in the penalty effect as the solution converged. The scheduling of the penalty was based on a previous work (*60*).

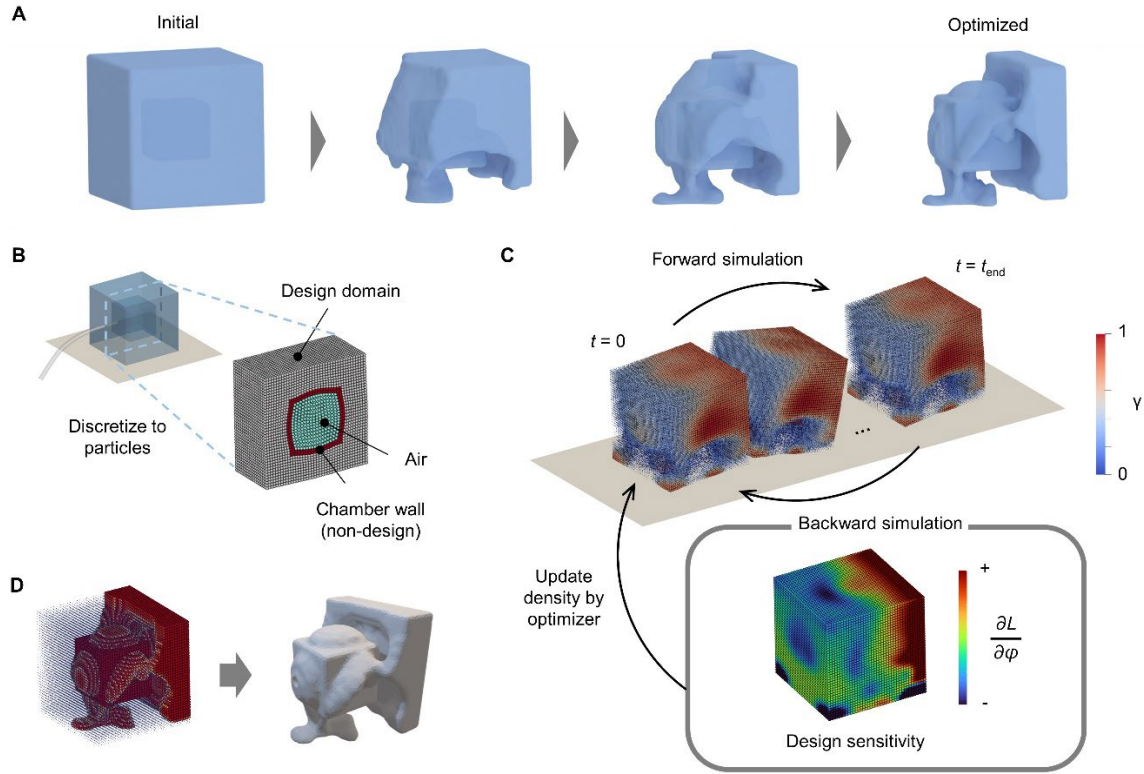

**Fig. 1. Shape synthesis of locomotive SoRos by topology optimization.** (**A**) Shape of a pneumatically actuated SoRo is synthesized by topology optimization, which optimizes the material distribution in the fixed design domain. This process looks similar to the evolution of biological organisms, yet it utilizes gradient information to improve efficiency. The initial structure is a uniform material distribution with an air chamber. The void regions are represented by material with zero density. (**B**) The design domain was discretized to particles for MPM simulation. The pneumatic actuator consists of chamber wall and air. (**C**) The MPM simulation was performed for the current design (showing the case of second from the left in (**A**)). The design sensitivity was derived via a backward simulation, tracking the dependency in reverse. The density distribution was updated using the design sensitivity. (**D**) The optimized result was postprocessed. The interface between void and solid was extracted and the air hole at the rear end was added to facilitate the insertion of a tube for air input during actual 3D printing.

**Walker**

Using the above design process and 3D printing, we present *walker*, which is an optimized SoRo that performs horizontal locomotion on flat ground. **Figure 2** shows the optimization process and the movement of the *walker*. We set the initial shape as a 35 mm cube of fictitious density $\gamma = 0.5$ with a 17.5 mm cubic air chamber. The wall thickness of the chamber was set to 1.75 mm. The SoRo was discretized into 64,000 particles in a 100×100×100 grid representing a 175 mm cubic analysis domain. The grid was stationary in space, and it had been set to cover the entire movement range of the robot.

The material properties of the soft body were set for the 3D printed soft material with equilibrium Young's modulus $E_0 = 0.44$ MPa, Poisson's ratio $\nu = 0.4$, and density $\rho = 1.07 \times 10^3$ kg/m$^3$. Further details on the material are presented in the "Soft robot fabrication" section. The material was expected to be nearly incompressible with a Poisson's ratio of 0.5. However, as we used the MPM formulation assuming compressibility, the Poisson's ratio $\nu$ was set to 0.4 to avoid numerical instability. The details regarding the numerical stability and viscosity parameters are discussed in the "Viscoelastic parameter estimation" section. The physical properties of the air were set as bulk modulus $k = 0.14$ MPa and shear viscosity $\mu_v = 1.83 \times 10^{-5}$ Pa s. The volumetric viscosity $\zeta$ was set to zero. The density of air was artificially set to $1.00 \times 10^2$ kg/m$^3$ to satisfy the Courant-Friedrichs-Lewy (CFL) condition (*73*) regarding the propagation speed of the pressure wave. The effect on the locomotion under this condition was small because the air mass in the chamber was one-tenth that of the chamber wall.

The pressure waveform used for the actuation in the simulation was measured while the valve was switched between atmospheric pressure and 80 kPa at 5 Hz with a 1:1 duty ratio using a SoRo prototype, as shown in Fig. S1. The reason we used the experimental data instead of a simplified waveform is that the pressure rise rate greatly influences the robot's movement. Moreover, modeling pressure changes involves many factors, such as the robot's stiffness during inflation and the properties of the air tubing. Using the experimental data allows us to fully reproduce the real-world behavior without complicating the simulation model. We measured the pressure using a pressure sensor TBPDANN150PGUCV (Honeywell Sensing and Productivity Solutions, Charlotte, NC, USA) equipped with a data acquisition device USB-1208FS-Plus (Measurement Computing Corporation, Norton, MA, USA). The pressure waveform is shown in Fig. S2. The actuation was started at $t_{\text{start}} = 0.25$ s to avoid exploiting the bounce by self-weight for locomotion. The actuation was applied for 0.5 s, therefore the simulation ended at $t_{\text{end}} = 0.75$ s. The time step of the MPM simulation was set to $10^{-5}$ s. For the topology optimization parameters, we set the filter radius as 0.00525 and the β in the projection as 8. The learning rate of the Adam optimizer was set to 0.02. To enforce left/right symmetry, the design sensitivity was averaged on the left/right side of travel direction in each optimization iteration. The shape should ideally remain symmetric even without this operation because the problem setting is symmetric. However, due to numerical errors from GPU parallelization, symmetry was explicitly enforced. We ended the optimization when the rate of change in the averaged objective function was less than 0.1%, based on the last four iterations and the four iterations before that.

The shapes and states at $t_{\text{end}}$ for several optimization iterations are shown in **Fig. 2A**. Movie S1 shows the shape synthesis process and locomotion behavior over several iterations. As shown in **Fig. 2A**, the initial structure could not locomote because it inflated and deflated uniformly. As the optimization proceeded, leg-like structures were formed on the front side (front leg) and underside of the chamber (back leg). **Figure 2B** shows the optimization history of locomotion distance $\mathcal{L}$ and constraint function $C$ that penalizes the intermediate density. During the first 50 iterations, the locomotion distance increased rapidly. This is because the design variable $\phi$ in each particle reaches from 0 (middle of void and material) to −1 (void) or 1 (material) by 50 updates with a learning rate of 0.02 if the sign of design sensitivity is always positive or always negative through these iterations. Subsequently, the locomotion distance gradually improved as more detailed structures emerged. In the second half of the optimization (150–339th iterations), the change in locomotion distance was small, and updates were mainly for satisfying the constraint of Eq. 16b. Using the projection scheme and constraint function, the optimized result consisted mostly of solids and voids, with few intermediate states. The optimization history of the robot's center of gravity and total mass is shown in Fig. S3 (for further details, see Section S1). Note that the optimization result is reproducible because the gradient-based method is deterministic, yielding identical results with each run. The average computation time for a single iteration was 413.6 s, consisting of 70.3 s for the forward simulation and 333.4 s for the backward simulation. The total time for 339 iterations was 140,210 s (38.9 hours). The total computational workload required for the entire optimization was equivalent to approximately 1946 times that of the forward simulation, which is substantially less compared to the case with EAs (*30, 31, 34*).

**Figure 2C** shows the movement of *walker* during the simulation (Movie S2 presents the video). When the chamber pressure was low (0 s), the front and back legs were closed, and the sole of the back leg was parallel to the ground. By increasing the pressure, the chamber was inflated such that the back leg kicked the ground (0.03 s). These leg movements are achieved using a SoRo shape. The top and back of the chamber were covered with material to maximize the inflation of the front and bottom of the chamber. In the fully inflated state (0.10 s), the front and back legs landed the farthest apart. During the deflation (0.13 s), the *walker* leaned forward to attract the back leg. After deflation (0.2 s), the posture returned to its initial state (0 s). Therefore, the movement was repeated periodically to facilitate stable forward movement. The effect of Young's modulus and pressure settings is discussed in Section S2.

The optimized structure was fabricated using digital light processing (DLP) 3D printing (further details in "Soft robot fabrication" section). Air pressure was controlled using a digital dispenser DC200 (Fisnar, Germantown, WI, USA). A tube was attached to the hole at the back of *walker*. The preliminary study showed that 80 kPa, the same as the pressure waveform measured for the simulation, resulted in overinflation, and it was impossible to exhaust air during one cycle. Therefore, the pressure used in the experiment was set to 50 kPa. The pressure frequency was set to 5 Hz. **Figure 2D** shows the movement of fabricated *walker* during the experiment. The behavior of *walker* was similar in both the experiment and simulation. The stride in one cycle was comparable to that in the simulation. Movie S2 shows locomotion in both simulation and experiment (for further detailed comparison, see "Simulation-to-real gap" section).

**Figure 2E** shows the trajectory of *walker* when operated for an extended period (Movie S3 presents the video). To verify the robustness of the optimized design, the *walker* was made to walk uphill at 10%, 20%, and 30%. On flat ground, *walker* continued to step forward repeatedly without losing balance. The locomotion speed was 53.5 mm/s, which is 1.7 Body Length/s (BL/s). To confirm superiority of the optimization results, performance of designs during the optimization process were investigated by experiments (for details of experiments, see Section S3). The experimental performance showed the same trend as the calculations, i.e., the locomotion speeds at the 20th and 100th iterations were 6.42 and 22.4 mm/s, respectively (Fig. S6, movie S4). We compared the locomotion speed with previous works which uses tethered pneumatic SoRos for a walking task on flat ground. The locomotion speeds are 33.3 mm/s (0.128 BL/s) in (*74*), 12.4 mm/s (0.0354 BL/s) in (*75*), 16.3 mm/s (0.120 BL/s) in (*4*), 25 mm/s (0.18 BL/s) in (*76*), and 25.6 mm/s (0.180 BL/s) in (*5*). As shown in **Fig. 2E**, the *walker* could walk at a 20% incline. Finally, at a slope of 30%, it became difficult to kick out and the progress was noticeably shorter. The locomotion speeds at 10%, 20%, and 30% inclines were 37.0, 32.5, and 9.7 mm/s, respectively.

To further analyze the changes in optimal structures with respect to task/environment differences, optimization calculations were performed for walking on different angles of incline (**Fig. 3**) in the same setting as the flat case. As depicted in **Fig. 3A**, the optimized result for the 30% incline revealed a structure with front and back leg-like formations similar to the original walker, but more complex. The legs were split into left and right, increasing the contact area with the ground and improving traction. An almost disconnected structure was observed at the upper front end of the body, likely for weight balance. The back legs consisted of two thin members, increasing compliance compared to the original walker. In the simulation (**Fig. 3B**, movie S5), it locomoted 18.2 mm in 0.5 s.

**Figure 3C** is the optimization result on a 57.7% (30-degree) slope, showing thinner and more complex shape. It is noteworthy that features of the optimized results gradually changed with respect to the environment, i.e., the slope angle in this case. Specifically, the structure of the back legs drove from the bottom of chamber in the flat case (**Fig. 2**), from the bottom and side in 30% (**Fig. 3A**), and from two places on the side in 57.7% (**Fig. 3C**). In the simulation, it could walk on the steep slope of 57.7% for 14.9 mm in 0.5 s. Note that long time simulations with MPM deviated the particle arrangement from the initial state. As a result, the locomotion on 57.7% slope in **Fig. 3D** slowed down in 1 to 1.5 s, as shown in movie S5.

These results demonstrated that the proposed method can be used to design a SoRo shape that functions in the real world without assuming the features of a soft body, such as arms and legs. In addition, it has been demonstrated that the proposed method provides insights into the morphology of SoRos depending on the task and environment.

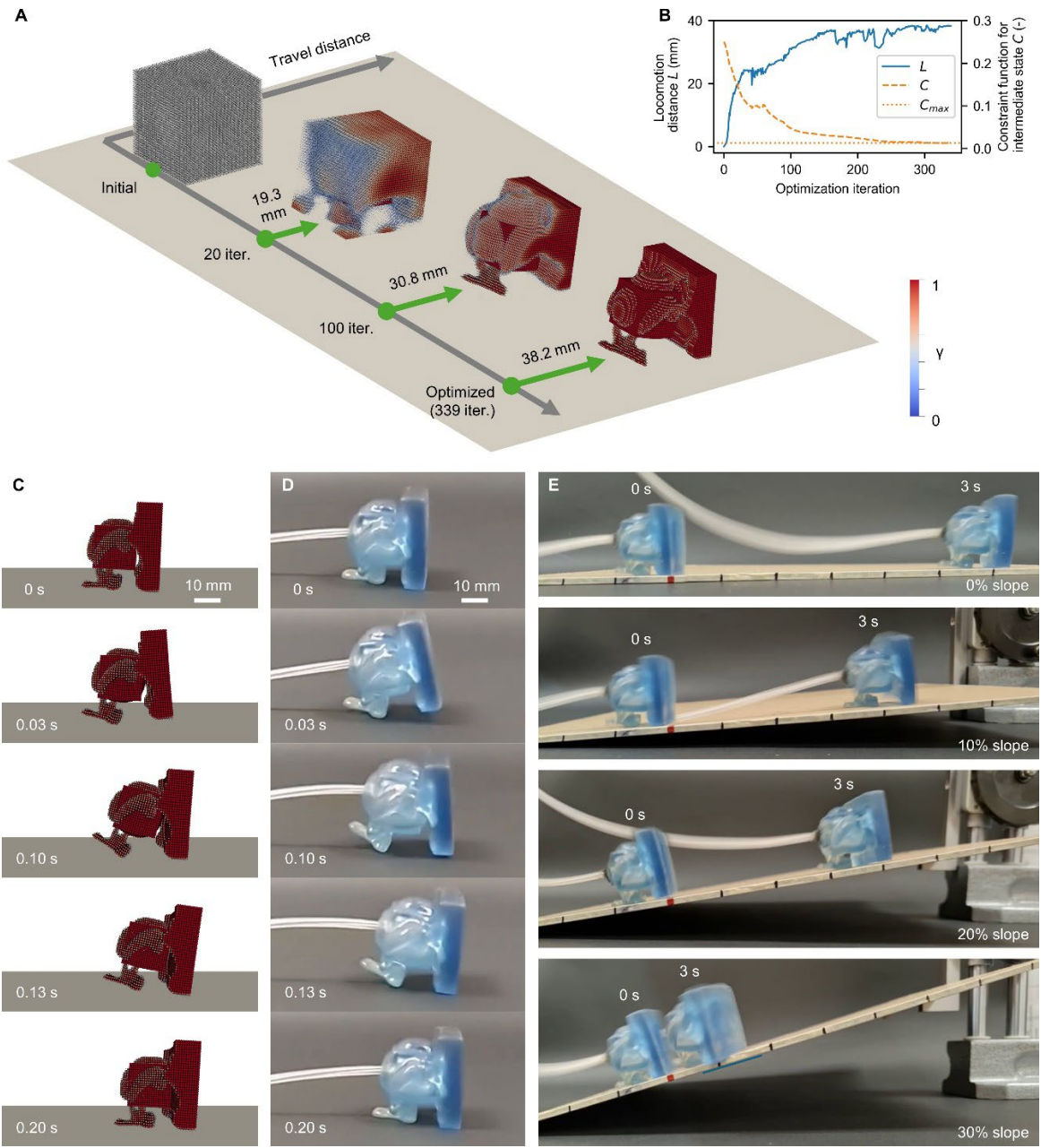

**Fig. 2. Optimized walker.** (**A**) The body shape of the SoRo was synthesized through the iterative optimization process. The state at the end of the simulation (0.75 s) is shown. Leg-like structure was formed, and the intermediate density was eliminated in the optimized structure (movie S1). (**B**) The locomotion distance $\mathcal{L}$ was improved by the optimization, and the constraint function $C$ was satisfied in the optimized result. (**C**) In the simulation, the optimized SoRo stepped forward by inflating the bottom and front face of the air chamber (movie S2). (**D**) The experiment shows the similar movement to the simulation (movie S2). (**E**) The fabricated SoRo could locomote stably for an extended time. On the 10% and 20% slopes, which are not considered in the optimization, it could locomote without losing the balance (movie S3). The marker on the board represents 25.4 mm (1 inch).

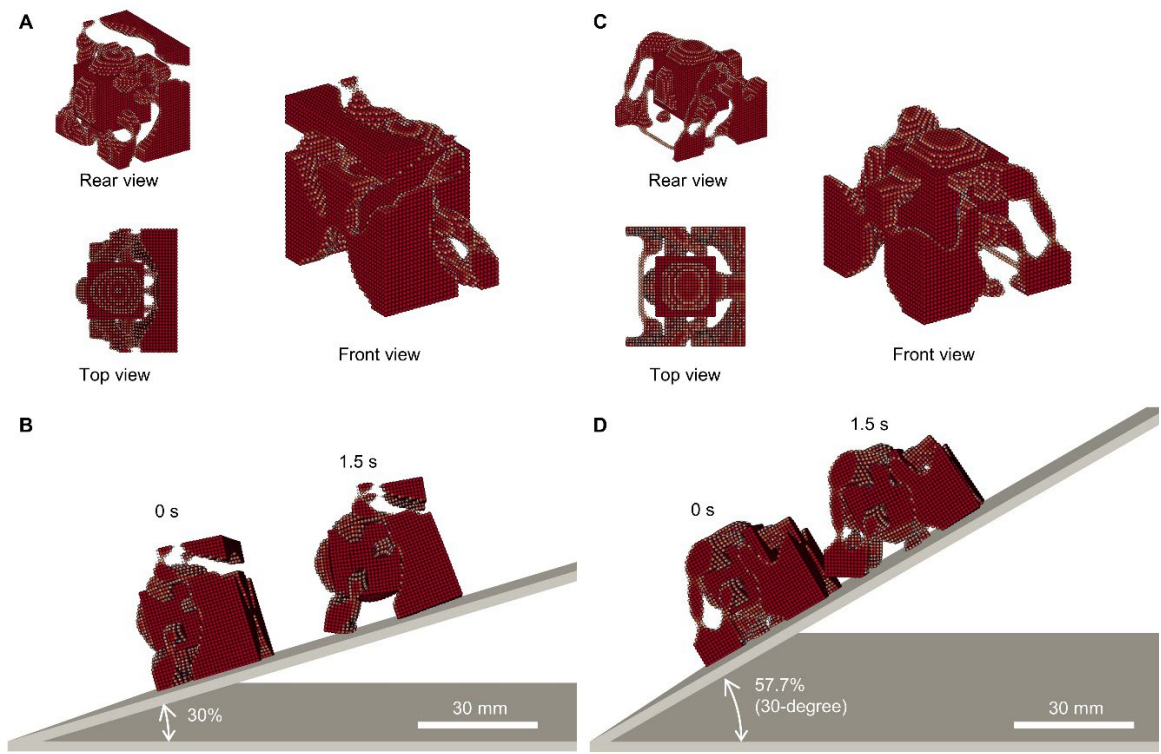

**Fig. 3. Optimized walking SoRo designs for slope. (A)** Optimized shape for 30% slope. The optimization process led to a structure where the contact area with the slope was divided into four leg-like parts to avoid losing traction. In addition, a center of gravity balancing structure emerged at the top of the front end with a very weak connection to the main body. **(B)** Locomotion on the 30% slope was similar to that on a flat ground. The optimization led to long contact parts at the front, which improved traction. **(C)** Optimized shape for 57.7% (30-degree) slope. The optimization led to a structure similar to the 30% slope case, where the contact area was divided into four leg-like parts, although the rear parts were connected with a thin structure. Small contact parts at the rear end of the front legs also emerged to prevent slipping. **(D)** In the simulation, the optimized robot could locomote on the 30-degree slope. The optimized result avoided placing objects at the top, allowing the robot to step forward without rolling backward.

**Climber**

Next, the proposed method was utilized for a more challenging task: vertical locomotion against gravity (climbing). We designed *climber*, which is an optimized SoRo for climbing between two parallel walls 35 mm apart (**Fig. 4**). The initial shape was a rectangular prism with a height, width, and depth of 40.25, 35, 35 mm, respectively. The air chamber had the same dimensions (a 17.5 mm cube with a 1.75 mm thick wall) as the *walker* and was located at the center of robot. The robot was discretized into 83,200 particles in a 100×100×100 grid representing a 175 mm cubic analysis domain. The material properties and pressure waveforms were identical to those used in the *walker* experiment. In contrast to *walker* optimization, actuation was initiated at the start of simulation ($t_{\text{start}} = 0$ s) to avoid falling. The actuation was applied for 0.5 s, therefore the simulation ended at $t_{\text{end}} = 0.5$ s. In addition, gravity was increased linearly from zero to 9.8 m/s² every 20 iterations until 200 iterations were reached to prevent falling during the early stages of optimization. The time step of MPM simulation was set to $10^{-5}$ s. For the topology optimization parameters, we set the filter radius to 0.004375 and the β in the projection to 8. The learning rate of Adam optimizer was set to 0.02. A symmetric condition was introduced by averaging the design sensitivity on the YZ and ZX planes when the normal direction of the wall was the X-axis and the direction of travel was the Z-axis. The computer was the same as that used in the *walker* optimization. The average computation time for a single iteration was 357.0 s, consisting of 56.0 s for the forward simulation and 286.8 s for the backward simulation. The total time for 373 iterations was 133,161 s (37.0 hours).

The shapes and states at $t_{\text{end}}$ for several optimization iterations are shown in **Fig. 4A**. Movie S6 shows the shape synthesis process and locomotion behavior over several iterations. As shown in **Fig. 4A**, the initial structure could not locomote because it was uniformly inflated and deflated. **Figure 4B** shows the optimization history of locomotion distance $\mathcal{L}$ and constraint function $C$ that penalizes the intermediate density. The locomotion distance improved mainly in the first 50 iterations. The subsequent 50–200 iterations were conducive to design updates toward a structure that supports the body against gravity, which increases every 20 iterations until it reaches 9.8 m/s². This can be seen from the locomotion distance, which decreased every 20 iterations and then recovered. After 200 iterations, the improvement in the locomotion distance was lower, and the design updates were mainly for satisfying the constraint of Eq. 16b. The optimization history of the robot's center of gravity and total mass is shown in Fig. S7 (for further details, see Section S4).

The optimized shape was more complex than that of the *walker*, as shown in **Fig. 4A**. The reason for this complex structure was that two types of motion are required for climbing without falling. The optimized climber had four arms on the upper side and two legs on the lower side. As shown in **Fig. 4C**, the two legs used a simple mechanism that kicked the wall by inflating the side of the chamber (0.10 s). However, the four arms on the upper side had a complex structure similar to a Y-shaped linkage mechanism. The four arms were pulled back when the chamber was inflated (0.10 s) to move forward, with the momentum of the kick from the legs on the lower side. The motion of *climber* was as follows: the upper four arms held the walls (0 s, 0.20 s), the lower two legs kicked the walls forward, and the upper four arms pulled back (0.10 s). Notably, such a movement was derived from optimization, starting from a uniform density field. Interestingly, the design of four arms and two legs was achieved using the optimization method without any human intervention.

The fabrication process and experimental setup for the air supply were identical to those used in the *walker* experiments. However, since the preliminary fabricated prototype exhibited a softer behavior than the simulation, we modified the surface data of the optimized result geometry, which was extracted as the isosurface of the density field with $\gamma = 0.5$. We increased the thickness by extruding the surface data in the normal direction by 0.4375 mm and fabricated this modified geometry for the experiment. The walls were made of glass and placed with a clearance adjustment to ensure contact between *climber* and the walls. As in the *walker* experiment, the pressure was set to 50 kPa. The frequency of the pressure was set to 3 Hz, although the optimization was performed at 5 Hz. The reason for this change was that preliminary experiments at 5 Hz showed asymmetric behavior that could not be recovered because of inconsistent contact with the walls (further details can be observed in movie S7). Although our method assumes a prescribed pressure waveform, pressure changes depend on the shape of the SoRo. The design dependency of pressure should be addressed in future studies.

**Figure 4D** presents the experimental results for the corresponding phases of the simulation. The fabricated *climber* exhibited a deformation similar to that in the simulation. The *climber* held the walls as expected and climbed them successfully. Movie S8 shows locomotion in both simulation and experiment (for further detailed comparison, see "Simulation-to-real gap" section). The snapshot of *climber* over an extended period is shown in **Fig. 4E**. *Climber* continued to climb step by step, with the speed of 7.7 mm/s, which is 0.21 BL/s. We compared the locomotion speed with previous works with similar problem setting, i.e., tethered pneumatic SoRos for climbing in vertically placed tubes. The locomotion speeds in previous works are 1.11 mm/s (0.014 BL/s) in 32 mm diameter (*77*) and 4.23 mm/s (0.045 BL/s) in 54 mm diameter (*7*). Note that the *climber* was sensitive to disturbances, that is, the initial placement inclination and left–right difference in the contact condition. As shown in movie S9, the *climber* leaned to the left and right while its posture was corrected as its arms touched the walls. This is because both gravity and the attached tube can affect climbing, which is more challenging than walking on the ground. Even with these challenges, the experiment demonstrates that the proposed method can derive a shape with complex features for vertical locomotion against gravity, and that it functions in not only the simulation but the real world.

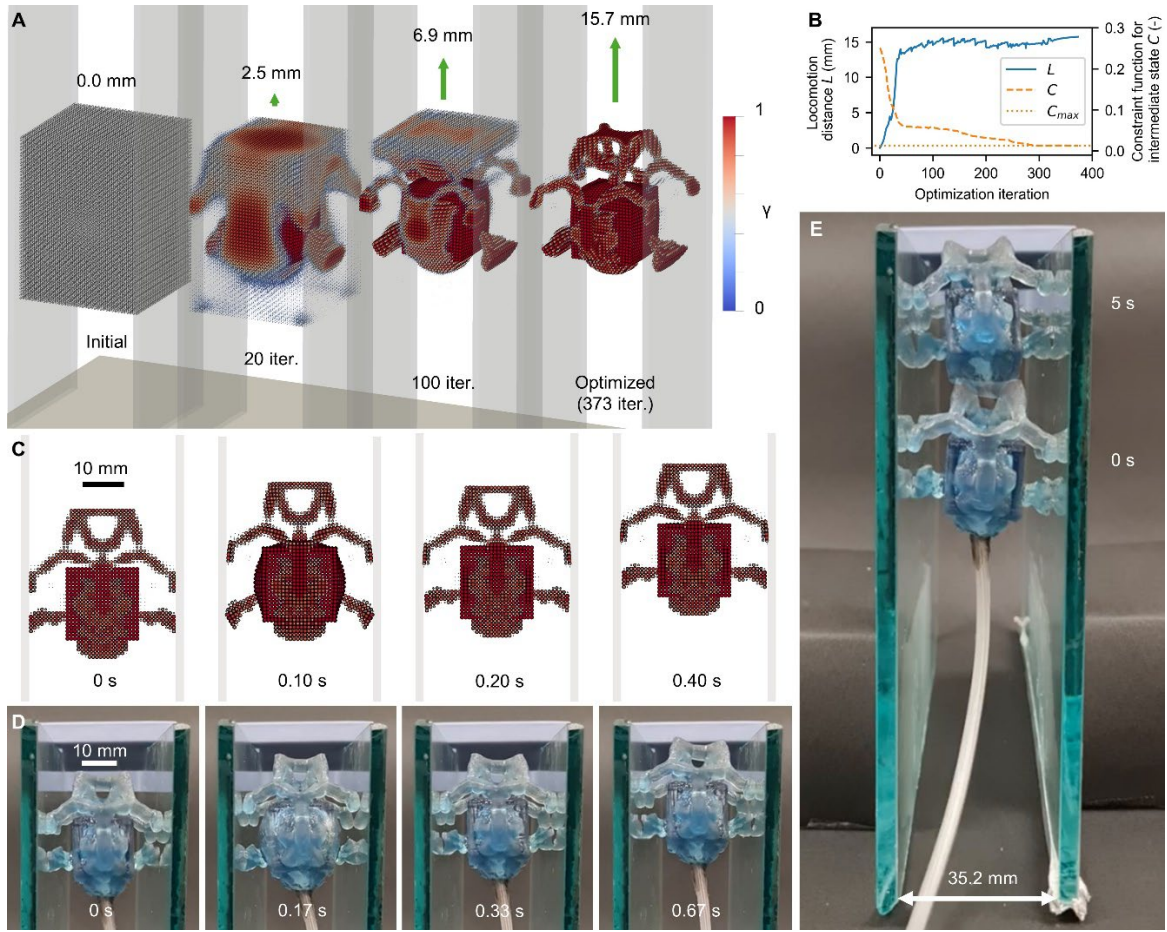

**Fig. 4. Optimized climber.** (**A**) The body shape of the climbing SoRo was synthesized through the iterative optimization process (movie S6). The state at the end of the simulation (0.5 s) is shown. The simulation for optimization was based on gradually increasing gravity; however, the visualization is based on a simulation with actual gravity for fair comparison. In the optimized result, four arms and two legs were formed on the upper and lower sides, respectively. (**B**) The locomotion distance $\mathcal{L}$ was improved by the optimization, and the constraint function $C$ was satisfied in the optimized result. The locomotion distance $\mathcal{L}$ shown here is based on increasing gravity depending on the iteration. (**C**) In the simulation, the optimized SoRo steps forward by kicking the wall with the lower legs and pulling back the upper arms (movie S8). (**D**) The experiment exhibited movement similar to that of the simulation (movie S8). (**E**) The fabricated SoRo can locomote continuously for an extended time (movie S9). It climbed 38.5 mm in 5 s.

**Simulation-to-real gap**

To analyze the gap between simulation and experiment in more detail, we tracked several parts of the SoRos and compared their trajectories. To track the experimental results from the video, we used Move-tr/2D (Library Inc., Tokyo, Japan), a software for analyzing the 2D behavior of objects from the time series of images.

For the *walker*, we tracked four points as shown in **Fig. 5A**. **Figure 5B** shows the trajectory of each point for 0.6 s. For the trajectory of the front leg, the leg lifted higher and landed further forward in the experiment than in the simulation. One of the reasons is a change in the center of gravity caused by the tube attached to the tail. The bounce after the front leg landed was smaller in the experiment than in the simulation, since we removed the 4th and 5th Maxwell elements for numerical stability (see "Viscoelastic parameter estimation" section). Due to the different behavior of the front leg, the tail also had a larger displacement in the Z-direction in the experiment than in the simulation. On the other hand, the trajectories of the rear leg (Rear1 and Rear2) were consistent between the simulation and the experiment. The displacement of Rear1 in the Z-direction was small, indicating that the robot performed the minimum necessary movement suitable for flat ground.

These results show that the locomotion mode derived by optimization can be realized in the real world, although the robot's deformation is slightly different between the simulation and experiment, mainly due to the attached tube and material viscosity.

For the *climber*, **Fig. 5**C defines the points we tracked, and **Fig. 5D** shows the displacement of each point over time in the simulation and experiment, respectively. Since the actuation frequency differs between the simulation and experiment (5 Hz vs. 3 Hz, see "Climber" section), the horizontal axes in **Fig. 5D** are scaled to synchronize the periods and phases. The trajectory is shown for 5 cycles.

The *climber* locomotion in the simulation exhibited a symmetrical movement with limited displacement along the X-axis due to the symmetric shape of the robot (**Fig. 5D**). The upper and lower limbs switched between supporting the body by contact with the walls and moving upward without contact, as shown by their Z-coordinates increasing alternatively. Similar trends in the X- and Z-coordinates between the simulation and experiment indicate that the fabricated *climber* exhibited movements identical to those of the simulation.

However, the X-axis movement of the fabricated climber showed left-right asymmetry, unlike the simulation. This difference is attributed to variations in initial position and dimensional errors during fabrication. Specifically, minor discrepancies in initial positions can result in imbalances between the left and right limbs. During the experiment, the upper limb advanced further than expected compared to the simulation (0–0.4 s), and the lower left limb contacted the wall unexpectedly (0.6–0.8 s). Even with these deviations from the designed movement, the upper right and lower left limbs behaved similarly to the simulation, owing to the compliance of the soft body.

The *climber* exhibited continuous locomotion throughout the experiment, despite the observed differences from the simulation. The fabricated *climber* demonstrated robustness by leveraging the body compliance.

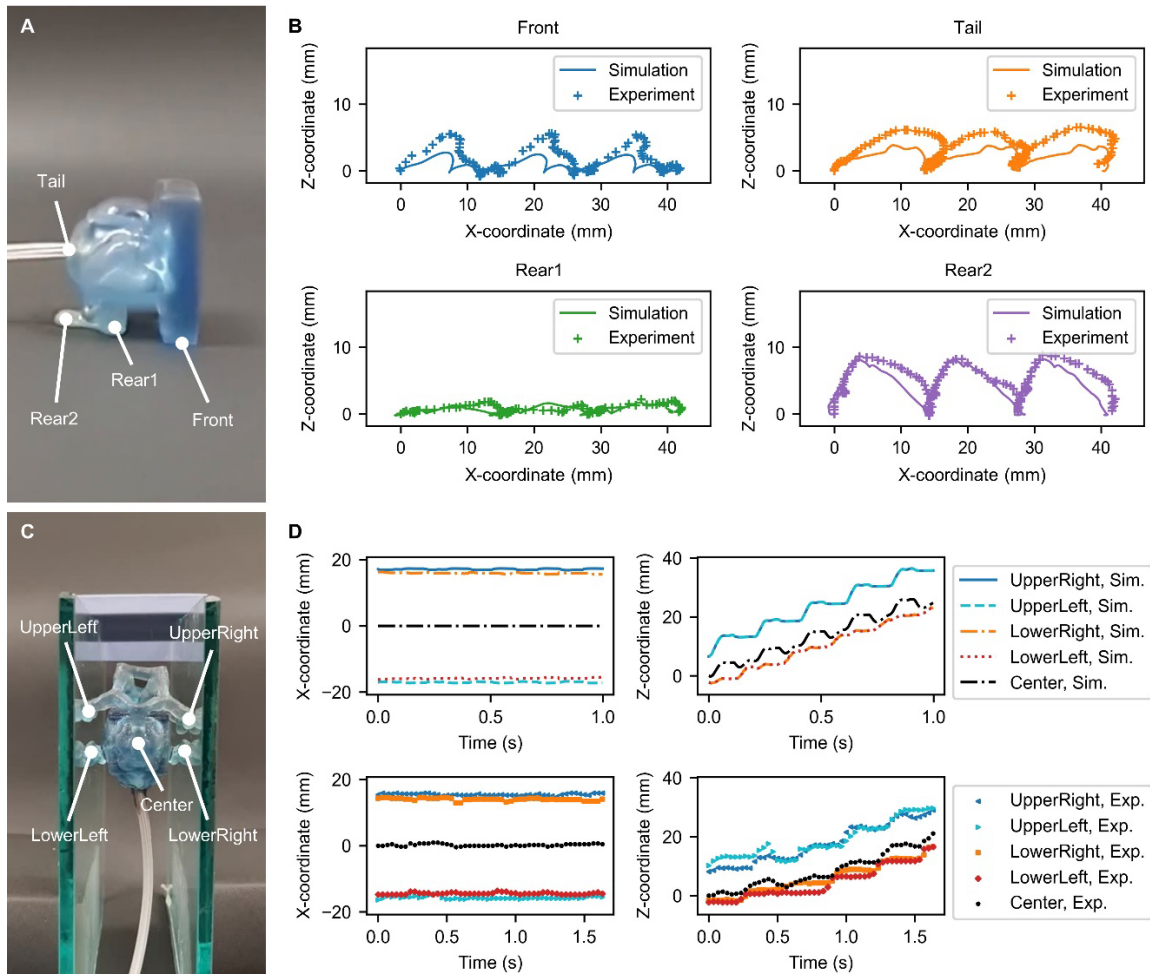

**Fig. 5. Comparison of robot trajectories between simulation and experiment.** (**A**) Four points were selected to track the *walker*. The X-axis is the walking direction, and the Y-axis is the vertical direction. (**B**) Trajectories were extracted from the simulation and experimental snapshots. The origin was set to the initial position of each tracking point. The simulation and experimental results showed similar locomotion, with the primary difference being the higher lift of the front leg in the experiment, while the trajectories of the rear legs were closely aligned. (**C**) Five points were selected to track the *climber*. The X-axis is the direction normal to the walls, and the Z-axis is the climbing direction. (**D**) Trajectories were extracted from the simulation and experiment. The origin was the initial position of "Center." The fabricated *climber* achieved continuous locomotion, although it exhibited left-right asymmetric motion which was caused by an asymmetric initial position and dimensional errors during fabrication.

**DISCUSSION**

The design of pneumatic SoRos for locomotion is a challenging task due to the complex physical phenomena of locomotion and the high degrees of freedom required for structural design and actuation control. Focusing on the structural design, we proposed a method for the shape synthesis of locomotive SoRos using gradient-based topology optimization. It was assumed that the SoRos were actuated pneumatically using an air chamber. The body shapes of the SoRos were optimized to maximize locomotion distance. Through experiments, we demonstrated that the proposed method can be used to design a *walker* and *climber* for horizontal and vertical locomotion, respectively.

The *walker* and *climber* results suggested several promising features for SoRo design. First, the proposed design method does not require prior knowledge regarding the locomotion of SoRos or biological systems. Our two designs began with simple and similar geometries (a cube and a rectangular prism with a central cavity). The optimization proceeded and generated limb-like structures as required, similar to evolution in nature, depending on the mode of locomotion. Second, the proposed method can provide feasible designs even for tasks or environments where we do not know how to accomplish the requirement. This has been demonstrated in the *climber* example, where the linkage-like structure was formed to achieve multiple deformation modes using only one actuator. Existing design methods are based on approaches that are built from the functions of primitive components or are inspired by the mechanisms of biological organisms. By contrast, our proposed method starts by formulating an objective function, guiding the design exploration. This ensures a goal-oriented approach without the need to predefine locomotion modes. Third, the dynamic behaviors of SoRos can be considered through simulations. The effects of inertia on the SoRo deformation and posture changes were difficult to predict before the experiment. The proposed method is particularly advantageous for large-scale robots and high-speed locomotion.

This study also suggests the directions for future research. One important issue is ensuring the robustness of the optimization results. The low resilience to posture changes in the *climber* experiment was caused by the difference between simulation and real world with disturbances. If robust optimization is realized, it can be applied to more practical engineering problems. Yet, a valid design solution may not be obtained for certain specific design problems. The reason is that gradient-based optimization may yield undesirable local solutions when there is no monotonically decreasing path to a satisfactory solution. For example, problems requiring assembly actions that involve complex manipulations are expected to be difficult. Additional research is required on formulating such problems such that optimization can proceed successfully. In addition, the design dependency of the actuation pressure should be addressed, as discussed in the "Climber" section.

Another direction is to extend the optimization to include shapes of the chamber wall and air region (**Fig. 1B**), which are fixed in this study. Furthermore, optimization could be extended to handle multiple pneumatic actuators and optimize their actuations in time series. If multiple actuators can be optimally placed and each can be driven optimally, it is possible to automatically design SoRos that can perform a wide variety of tasks. Optimization of the material layout, actuator layout, and time-varying actuation has recently been realized (*60*). Although promising results have been obtained, showing animal-like movements in calculations, a problem setup that can be manufactured has not yet been studied. Therefore, methods for problem definition and experimental verification to realize SoRo optimization, including actuators, are important for future research.

In summary, this study presented a framework for the computational synthesis of SoRos for a given locomotion task without a prior assumption of the shape. These results indicate that the proposed optimization framework can play a role in future SoRo designs that might one day approach the sophisticated shapes found in biological organisms.

## MATERIALS AND METHODS

### Material point method

MPM is a hybrid Eulerian-Lagrangian approach to simulate continuum material such as solids, liquids, and gases. Materials are represented by material points, where the governing equations are solved on a grid. The MPM procedure comprises three steps: particle-to-grid transfer, grid operation, and grid-to-particle transfer. The MPM can handle large deformations owing to particle representation; therefore, it is suitable for soft-body simulations. In addition, multi-material interactions can be treated, enabling the consideration of fluid–structure interactions. In this study, we employed MLS-MPM (*65*)) for computational efficiency and ease of implementation.

### Computational setup

All numerical simulations and optimizations were performed on a computer with an NVIDIA A100 80GB, two AMD EPYC 7452 32-Core Processors, and 512GB of memory. The program was implemented in Python 3.8.10, using Taichi 1.2.0. The initialization and optimization loops were calculated for the CPUs. The MPM simulation was performed using Taichi (*68*) for parallel computing on a GPU and automatic differentiation. A technique referred to as checkpointing (*60*, *66*) was used to fit the GPU memory limit. Checkpointing can reduce memory usage by recomputing the forward simulation steps in the backward simulation. This is advantageous when the number of time steps $N_t$ is large because memory usage can be $\mathcal{O}(\sqrt{N_t})$ whereas the increase in computational cost remains $\mathcal{O}(N_t)$.

### Soft robot fabrication

The resin used in the SoRos was a combination of butyl acrylate (BA) (Sigma-Aldrich), isodecyl acrylate (IDA) (Sigma-Aldrich), 2-Hydroxyethyl acrylate (2HEA) (Sigma-Aldrich), and aliphatic urethane diacrylate (AUD) (Ebecryl 8413, Allnex, GA, USA) at a weight ratio of 3:3:2:2. 819 Photoinitiator (Phenylbis (2,4,6-trimethylbenzoyl) phosphine oxide) (Sigma-Aldrich) and Clariant Solvaperm® Blue 2B-CN (light absorber) were added with 0.6 and 0.01 weight %, respectively.

An in-house DLP printer was used in this study. The printer comprised a projector PRO4500 (Wintech Digital, Carlsbad, CA, USA) and a *z*-motion stage LTS150 (Thorlabs Inc., Newton, NJ, USA).

### Dynamic mechanical analysis of soft material

The viscoelastic properties of the MPM simulations were measured by dynamic mechanical analysis (DMA). A dynamic mechanical analyzer Q800 (TA Instruments, New Castle, DE, USA) was used for the measurements. The test piece had a rectangular shape with a length, width, and thickness of 16, 4.1, and 0.6 mm, respectively. The testing condition was a tensile loading at frequencies 0.1, 0.5, 1, 2, 5, 10, and 20 rad/s at temperature range of -85 °C to 40 °C in 7 °C increments. The frequency response diagrams for frequencies less than $1 \times 10^7$ rad/s are shown in Fig. S8. The diagrams show the values of the storage modulus $G'$ and loss modulus $G''$ on the vertical axis. In this study, by leveraging the principle of time-temperature correspondence, the long-term behavior of soft materials was estimated using the time-temperature superposition method. The modulus curves were horizontally shifted towards a reference temperature (21.8°C) until an exact superposition was achieved.

### Viscoelastic parameter estimation

The storage modulus $G'$ and loss modulus $G''$ were represented as functions of frequency ω in the following equations (*78*):

$$G'(\omega) = g_\infty + \sum_i \frac{g_i \omega^2 \tau_i^2}{1 + \omega^2 \tau_i^2}, \tag{17a}$$

$$G''(\omega) = \sum_i \frac{g_i \omega \tau_i}{1 + \omega^2 \tau_i^2}. \tag{17b}$$

In this study, the number of terms was set to five. The relaxation time $\tau_i$ was put at equal intervals on a logarithmic scale, and unknown relative moduli $g_\infty$ and $g_i$ were approximated by fitting the Eqs. 17a and 17b to the experimental master curves of storage and loss modulus $F'$ and $F''$, respectively. Curve fitting was performed using a generalized reduced-gradient method to minimize the error norm $e$:

$$e = \sum_{k=1}^{M} \left[ \left( \frac{G'(\omega_k) - F'(\omega_k)}{F'(\omega_k)} \right)^2 + \left( \frac{G''(\omega_k) - F''(\omega_k)}{F''(\omega_k)} \right)^2 \right], \tag{17c}$$

where $M$ is number of sampling points and $\omega_k$ is sampling frequency. The abovementioned procedure was performed using commercial curve-fitting software, particularly designed for the Prony series (79). The obtained relative moduli $g_\infty$, $g_i$ and corresponding relaxation time $\tau_i$ in the Prony series are shown in Table S1.

Note that the frequency range used for the fitting was set to less than $1 \times 10^7$ rad/s because the high-frequency range cannot be represented by the MPM time step $1 \times 10^{-5}$ s. In addition, to satisfy the CFL condition in the MPM, only Maxwell elements with $i = 1, 2,$ and 3 were used, removing $i = 4$ and 5, which had large moduli. The master curves of $G'$ and $G''$ for the three Maxwell elements are shown in the Fig. S8. Although the experimental values differed, the overall trends were consistent. For further verification, we compared the behavior of a 15 mm diameter ball dropped vertically in the experiment and MPM simulation with and without viscosity. As shown in Fig. S9, the MPM simulation with three Maxwell elements exhibited a behavior similar to that of the experiment, whereas the MPM simulation without viscosity yielded repulsion coefficients that were different from the experiment.

**Supplementary Materials**
    Sections S1 to S4
    Figs. S1 to S9
    Table S1
    Data S1
    Movies S1 to S9

**Acknowledgments:**

**Funding:** This project was partially supported by Toyota Motor North America, Inc. H.J.Q. acknowledges the support from a US AFOSR grant (FA9550-20-1-0306; Dr. B.-L. "Les" Lee, Program Manager).

**Author contributions:**
Conceptualization: HK, HJQ, TN
Methodology: HK, FG, SMM, MT, YL, CY, YS, AK
Investigation: HK, FG, SMM, MT, YL
Visualization: HK, FG
Project administration: HJQ, TN
Supervision: HJQ, TN
Writing – original draft: HK, MT, HJQ, TN
Writing – review & editing: HK, MT, CY, YS, AK, HJQ, TN

**Competing interests:** Authors declare that they have no competing interests.

**Data and materials availability:** All data needed to evaluate the conclusion of this manuscript are included in the main text and Supplementary Material. The source code can be provided by corresponding author pending license review and a completed material transfer agreement. Requests for the source code should be submitted to: Tsuyoshi Nomura (nomu2@mosk.tytlabs.co.jp).

Supplementary Materials for
Computational synthesis of locomotive soft robots by topology optimization

Hiroki Kobayashi et al.
Corresponding author: Tsuyoshi Nomura, nomu2@mosk.tytlabs.co.jp

**The PDF file includes:**
Sections S1 to S4
Figs. S1 to S9
Table S1
Legend for data S1
Legends for movies S1 to S9

**Other Supplementary Material for this manuscript includes the following:**
Data S1
Movies S1 to S9

## S1 Mass distribution change in walker optimization

Fig. S3 shows the changes in the robot's center of gravity and mass during optimization. Fig. S3(a) shows the center of gravity history overlaid on the optimized shape image. Fig. S3(b) shows the plot of the center of gravity and the mass through the optimization iteration. The mass increased rapidly in the early stage of optimization and then decreased gradually in the latter half. In the initial iterations, the center of gravity was raised vertically because the lower structure was softened to move the leg-like structures. Then, it was lowered back by removing unnecessary material from the upper part. Furthermore, the center of gravity moves continuously forward throughout the optimization and remains steady at the rear end of the front leg, which is suitable for retracting the back leg after stepping out.

## S2 Effect of Young's modulus and pressure settings on locomotion

To investigate how Young's modulus and pressure setting affects the soft robot locomotion, we have performed simulations for double Young's modulus with original pressure, and double Young's modulus with double pressure, compared with original properties (Fig. S4). The shape of soft robot is the same as *walker*, to compare the effect of modulus and pressure only. In double Young's modulus with the original pressure, the stride of soft robot was shorter (7.49 mm) than that of the original modulus (9.64 mm). On the other hand, the stride became longer (12.5 mm) in double Young's modulus and double pressure.

However, the increased stride did not effectively improve the locomotion distance. As shown in Fig. S5, the locomotion distance, which is the center of gravity change in X-axis, is 41.2 mm in double modulus and double pressure case. Despite a 30% increase in stride length, the locomotion distance improvement remains less than 8%. This can be attributed to an increase in bounce and a deviation from the original walking behavior, as can be seen from the position of the center of gravity in the Z-axis direction. These discussions suggest the material coefficients and pressure have a substantial impact on the optimal design of soft robots, and optimal shapes of soft robots vary depending on the change of material modulus or pressure.

## S3 Experiment for intermediate iteration design in optimization

The intermediate designs at the 20th and 100th optimization iterations were fabricated with the same material and print setting as that for the optimized design. The 3D models were extracted from isosurfaces at 50% density ($\gamma = 0.5$). Note that the isolated part at the bottom in 20th iteration design, which is shown in Fig. 3A, has been removed for 3D printing. The pressure and its frequency are set to 50 kPa and 5 Hz with 50:50 duty ratio, same as the setting for the optimized design.

Fig. S6 shows the experimental results of the 20th and 100th iteration designs. Compared to the optimized design (Fig. 2E), the locomotion speeds of the intermediate designs were noticeably shorter, i.e., 6.42 mm/s at the 20th iteration design, and 22.4 mm/s at the 100th iteration design. The trend of locomotion distance is consistent with the increase of objective function as optimization proceeds. In the early stages of optimization, more material is used to make the structure viable. As a result, the strides are shorter due to smaller deformations. These results confirm that updating the design in optimization calculations contributes to improved performance in reality.

## S4 Mass distribution change in climber optimization

Fig. S7 shows the changes in the robot's center of gravity and mass over the whole iterations. Fig. S7(a) shows the center of gravity history overlaid on the optimized shape image. Fig. S7(b) shows the plot of the center of gravity and the mass through the optimization iteration. Similar to the *walker*, the mass of the *climber* increased early in the optimization and then decreased. The mass of the final design is lower than that of *walker* (Fig. S3) because *climber* locomotes against gravity. The center of gravity moved to the front part, which required the complex action of extending the arms forward and then holding the arms against the wall, but as the wasted material was removed, it returned to near the center of the chamber.

Table S1. Estimated relative moduli $g_\infty$, $g_i$ and corresponding relaxation time $\tau_i$ in Prony series.

| Number of Maxwell elements | $g$ (-) | $\tau$ (s) |
|---|---|---|
| $\infty$ | $9.06 \times 10^{-4}$ | - |
| 1 | $6.36 \times 10^{-4}$ | $2.73 \times 10^{-1}$ |
| 2 | $2.09 \times 10^{-3}$ | $7.56 \times 10^{-3}$ |
| 3 | $1.27 \times 10^{-2}$ | $2.09 \times 10^{-4}$ |
| 4 | $1.25 \times 10^{-1}$ | $5.77 \times 10^{-6}$ |
| 5 | $8.59 \times 10^{-1}$ | $1.59 \times 10^{-7}$ |

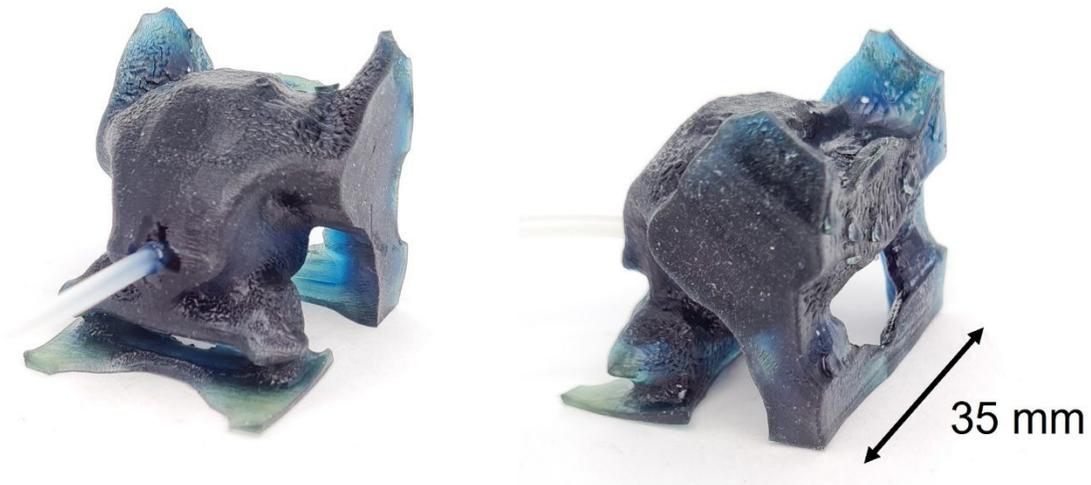

**Fig. S1. Prototype of soft robot used for the air pressure measurement.** This prototype consists of the materials described in "Soft robot fabrication" section.

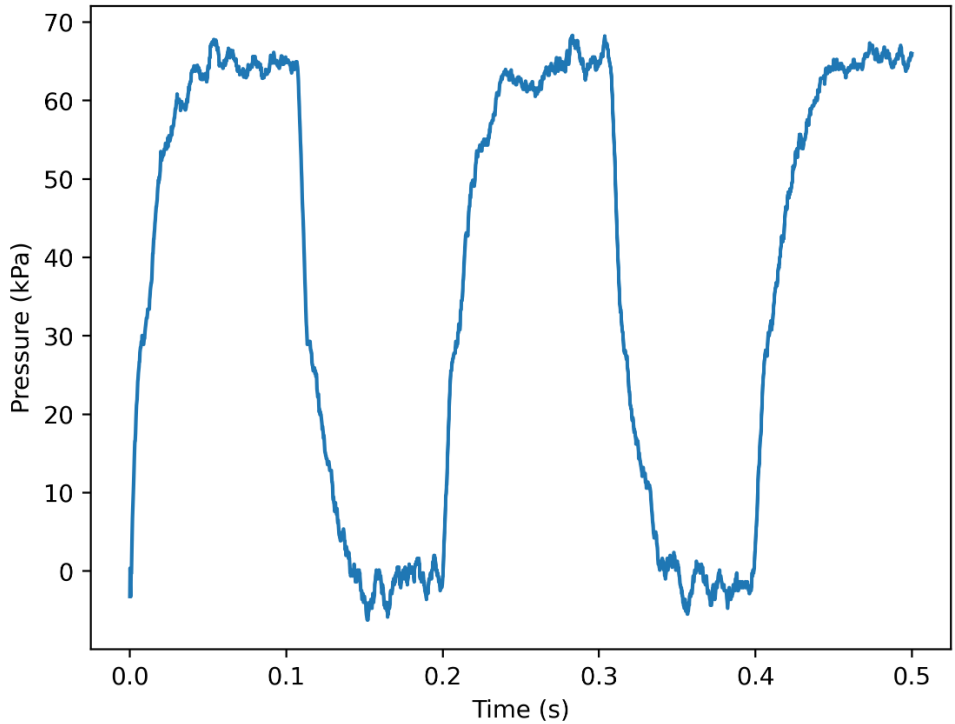

**Fig. S2. Waveform of the pressure used for the simulation.** The pressure was measured for the prototype shown in Fig. S1. The data is available in Data S1.

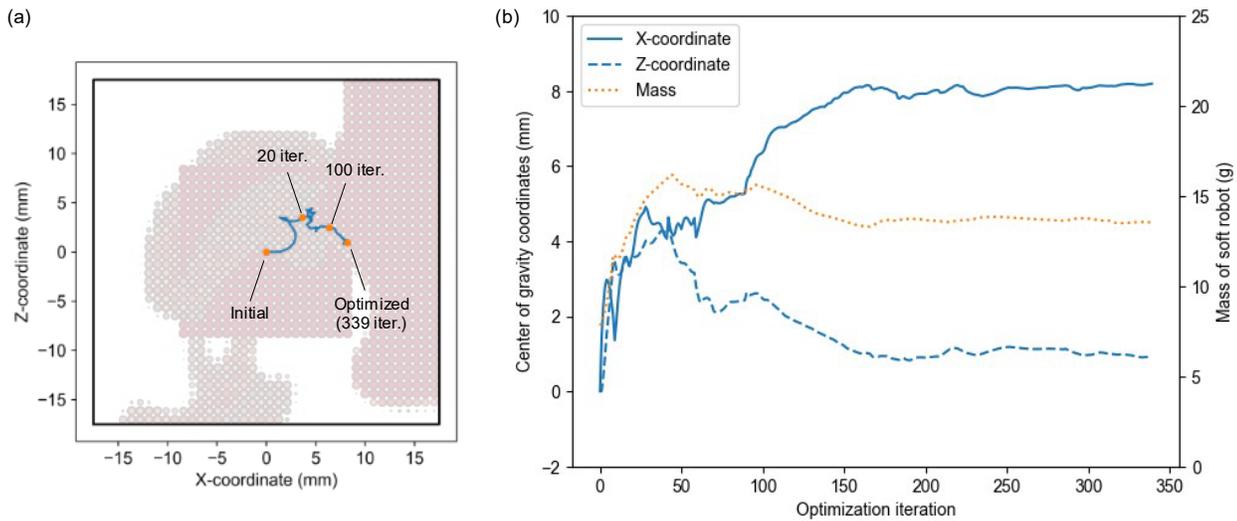

**Fig. S3. Center of gravity coordinates in static state and mass of the robot in *walker* optimization.** (a) The center of gravity coordinate was changed through the design updates. The plot is overlaid on the optimized shape image. The coordinate origin is defined as the center of pneumatic actuator. X-axis and Z-axis are walking direction and vertical direction, respectively. (b) Detailed plot of the robot's center of gravity and mass. The center of gravity moved forward and slightly upward as the optimization proceeded. The mass was increased in early stage of optimization but decreased in the later iterations.

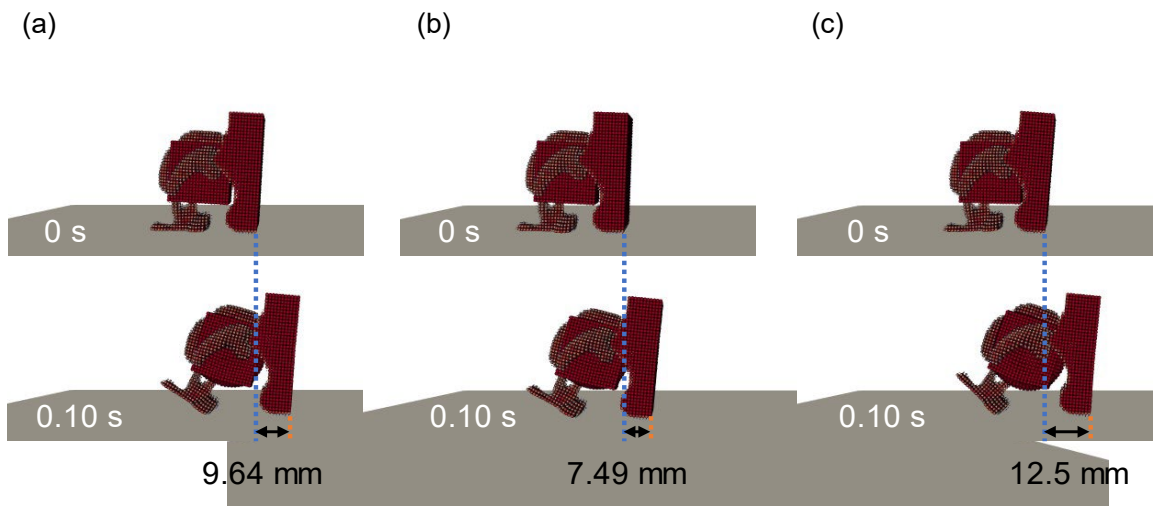

**Fig. S4. Stride distances for different Young's modulus and different pressure settings.** (a) original modulus and pressure. (b) double modulus with original pressure. (c) double modulus with double pressure.

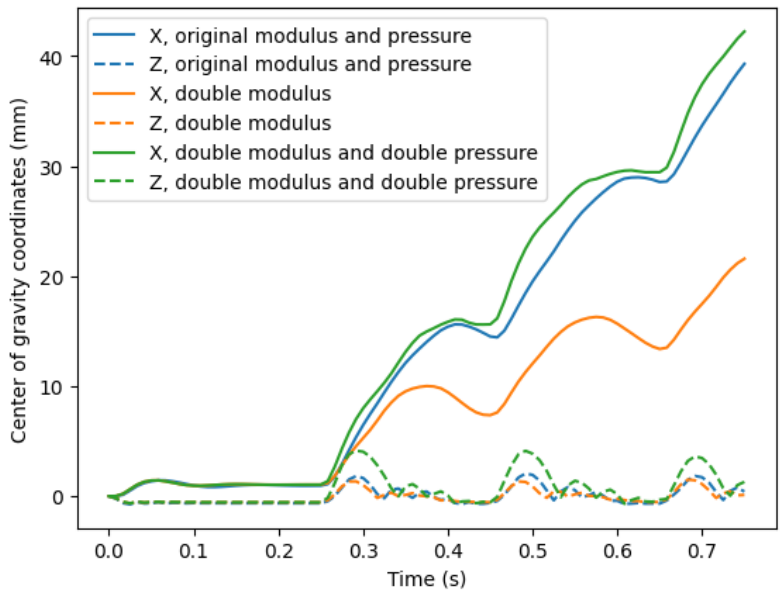

**Fig. S5. Center of gravity plot for different Young's modulus and different pressure settings for *walker*.** X-axis is the walking direction (horizontal), and Z-axis is the vertical direction. The origin of center of gravity is based on initial state (0 s).

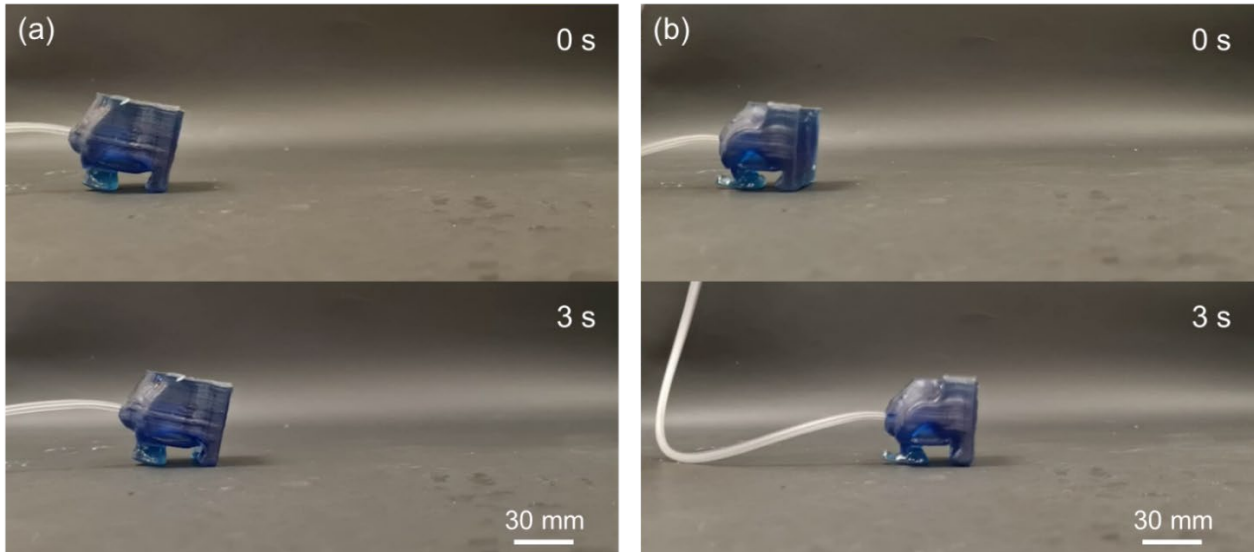

**Fig. S6. Experimental results of intermediate design for walking task in 20th and 100th optimization iteration.** The soft robots are fabricated based on isosurfaces of 50% density ($\gamma$ = 0.5). (a) Snapshots of locomotion in 20th iteration design. The locomotion speed was 6.42 mm/s. (b) Snapshots of locomotion in 100th iteration design. The locomotion speed was 22.4 mm/s.

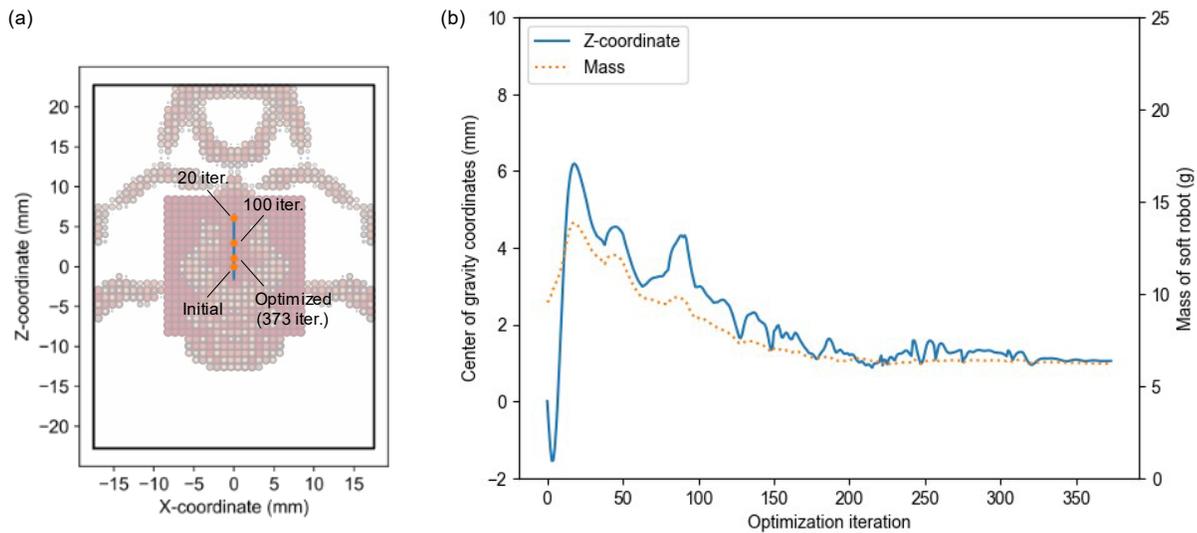

**Fig. S7. Center of gravity coordinates in static state and mass of soft robots in the *climber* optimization.** (a) The center of gravity coordinate was changed through the design updates. The plot is overlaid on the optimized shape image. The coordinate origin is defined as the center of pneumatic actuator. Z-axis is climbing direction. (b) Detailed plot of the robot's center of gravity and mass. The center of gravity moved upward. The mass was increased in early stage of optimization but decreased in the later iterations.

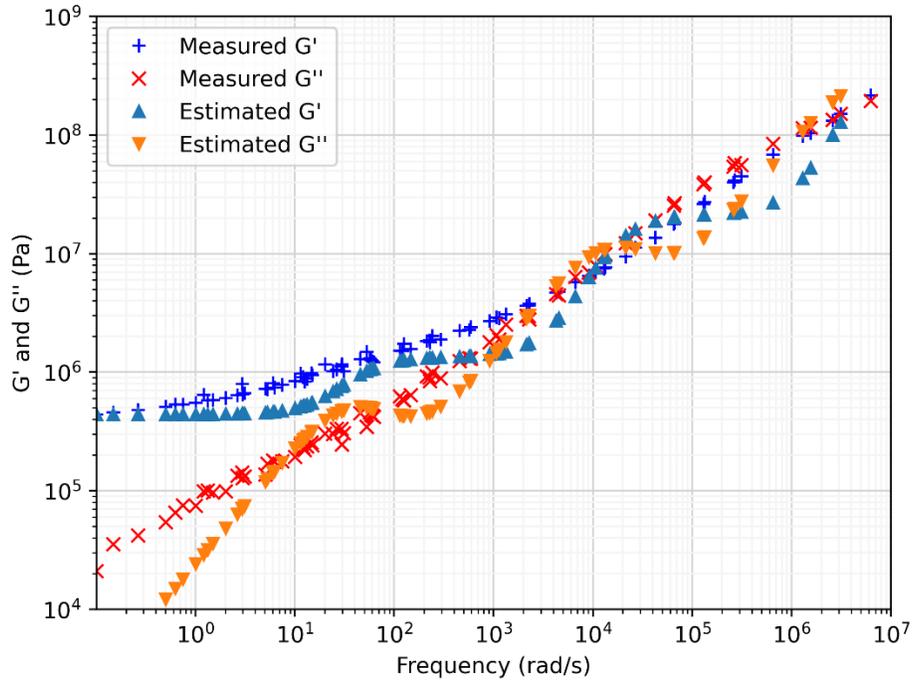

**Fig. S8. Frequency response diagrams of measurement results by DMA and estimation results using three Maxwell elements.** *G'* and *G"* represent the storage modulus and the loss modulus, respectively.

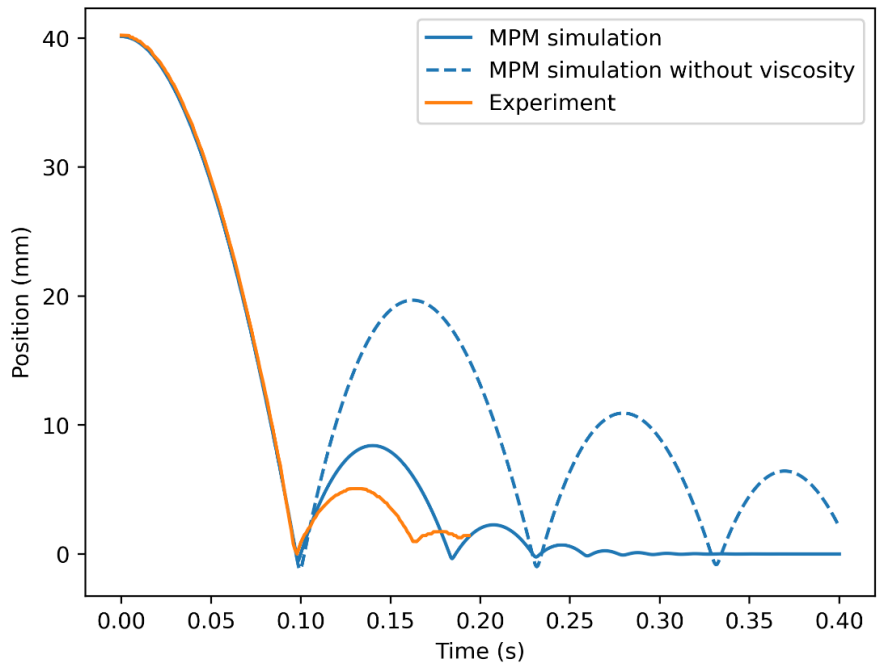

**Fig. S9. Time-series coordinate in vertical direction for the ball drop test.** To verify the effect of introducing viscosity, the MPM simulation result without viscosity are also shown.

**Legend for Data:**
Data S1.
Pressure waveform measurement data used for simulation.

**Legends for Supplementary Movies:**
Movie S1.
Shape synthesis process of *walker* and locomotion behavior over several optimization iterations.
Movie S2.
Comparison of *walker* locomotion between MPM simulation and experiment.
Movie S3.
*Walker* locomotion at 0%, 10%, 20%, and 30% inclines.
Movie S4
Locomotion of 20th and 100th iteration design for walking task.
Movie S5.
Optimized results in walking task on 30% and 57.7% (30-degree) inclines.
Movie S6.
Shape synthesis process of *climber* and locomotion behavior over several optimization iterations.
Movie S7.
*Climber* experiment result at 5 Hz.
Movie S8.
Comparison of *climber* locomotion between MPM simulation and experiment.
Movie S9.
*Climber* experiment for an extended time.